\documentclass[jkps,preprint,fleqn,showpacs,showkeys]{revtex4}
\usepackage{graphicx}
\usepackage{amssymb}
\usepackage{amsmath}
\usepackage{bm}
\usepackage{color}

\begin{document}

\def\cred{\color{red}}
\def\cblue{\color{blue}}

\setcounter{page}{0}


\title[]
{Soft X-ray Absorption Spectroscopy Study of Multiferroic
	Bi-substituted Ba$_{1-x}$Bi$_x$Ti$_{0.9}$Fe$_{0.1}$O$_3$}
\author{Hyun Woo \surname{Kim}}
\author{D. H. \surname{Kim}}
\author{Eunsook \surname{Lee}}
\author{Seungho \surname{Seong}}
\affiliation{
        Department of Physics, The Catholic University of Korea,
        Bucheon 420-743, Korea}
\author{Deok Hyeon \surname{Kim}}
\author{B. W. \surname{Lee}}
\affiliation{
        Department of Physics, Hankuk University of Foreign Studies,
        Yongin 449-791, Korea}
\author{Y. \surname{Ko}}
\author{J.-Y. \surname{Kim}}
\affiliation{
        Pohang Accelerator Laboratory, POSTECH,
        Pohang 790-784, Korea}
\author{J.-S. \surname{Kang}}
\email{kangjs@catholic.ac.kr}
\thanks{Fax: +82-2-2164-4764}
\affiliation{
        Department of Physics, The Catholic University of Korea,
        Bucheon-si 420-743, Korea}

\date[]{Received }

\begin{abstract}

The electronic structures of multiferroic oxides of
Ba$_{1-x}$Bi$_x$Ti$_{0.9}$Fe$_{0.1}$O$_3$ ($0 \le x \le 0.12$) 
have been investigated by employing photoemission spectroscopy 
and soft x-ray absorption spectroscopy (XAS). 
The measured Fe and Ti $2p$ XAS spectra show that 
Ti ions are in the Ti$^{4+}$ states for all $x$ and that
Fe ions are Fe$^{2+}$-Fe$^{3+}$ mixed-valent for $x>0$.
The valence states of Fe ions are found to be nearly trivalent 
for $x$=0, and decreases with increasing $x$
from being nearly trivalent ($v$(Fe)$\sim 3$) for $x$=0
to $v$(Fe)$\sim 2.6$ for $x$=0.12. The valence states of
both Ti and Ba ions do not change for all $x \le 0.12$.
Based on the obtained valence states of Fe ions,
the electronic and magnetic properties of 
Ba$_{1-x}$Bi$_x$Ti$_{0.9}$Fe$_{0.1}$O$_3$ are explored.

\end{abstract}




\maketitle

\section{INTRODUCTION}

Multiferroic materials exhibit the co-existence of multiple 
ordering, such as ferroelectricity, ferromagnetism, 
and ferroelasticity.\cite{Tokura10} 
Multiferroic properties observed in manganese (Mn)-based 
oxides\cite{Kimura03} and iron (Fe)-based 
oxides\cite{Ikeda05} have stimulated much research on 
multiferroic transition-metal oxides not only
due to the potential technical application but also due to
the scientific interest. 
As an attempt to achieve both ferroelectricity and ferromagnetism, 
Fe-doped BaTi$_{1-x}$Fe$_{x}$O$_3$ has been 
investigated by substituting ferromagnetic Fe ions 
in hexagonal ferroelectric 
BaTiO$_3$.\cite{Rajamani05,Ray08,Qiu10,Dang11,Chak11} 
Room-temperature ferromagnetism has been observed in 
BaTi$_{1-x}$Fe$_{x}$O$_3$ ($x<0.1$),\cite{Rajamani05,Ray08} 
and both tetragonal and hexagonal phases exist in 
BaTi$_{1-x}$Fe$_{x}$O$_3$ ceramics.
The saturation magnetization and the magnetic coercivity 
were found to depend on the doping level as well as 
the fraction of the hexagonal phase in the ceramics.\cite{Qiu10}.
In BaTi$_{1-x}$Fe$_{x}$O$_3$ system, oxygen vacancy defects 
and the valence states of the substituted Fe ions are 
expected to be important in determining the magnetic
and ferroelectric properties of the system.\cite{Chak11}
However, the reproducibility of multiferroicity and 
the effect of the oxygen vacancy in BaTi$_{1-x}$Fe$_{x}$O$_3$
need be investigated more systematically. 

Despite extensive studies on multiferroic transition-metal
(TM) oxides, the origin of the multiferroicity has not been
well understood yet.
Based on the XRD (X-ray diffraction) analysis of
BaTi$_{1-x}$Fe$_{x}$O$_3$,\cite{Dang11} the coexistence 
of Fe$^{3+}$ and Fe$^{4+}$ ions was suggested, implying
the availablity of the exchange interactions among 
Fe$^{3+}$-Fe$^{4+}$, Fe$^{3+}$-Fe$^{3+}$, 
and Fe$^{4+}$-Fe$^{4+}$. 
Further, the Fe $K$-edge XANES (X-ray absorption near-edge 
structures) study for BaTi$_{1-x}$Fe$_{x}$O$_3$\cite{Nguyen11}
have shown that both Fe$^{3+}$ and Fe$^{4+}$ ions exist 
in BaTi$_{1-x}$Fe$_{x}$O$_3$, with the maximum valence states 
at $x \sim 0.12$.
But XRD is not an element-specific experimental method 
that can determine the valence states of the constituent 
ions directly. On the other hand, Fe $K$-edge XANES  
arises mainly from the Fe $1s \rightarrow 4p$ absorption, 
and so Fe $K$-edge XANES does not directly represent
the Fe $3d$ configuration in the ground state of Fe ions.
Recently, the effect of the simultaneous substitutions of 
trivalent Bi$^{3+}$ ions for divalent Ba$^{2+}$ ions and 
trivalent Fe$^{3+}$ (or tetravalent Fe$^{4+}$) ions 
for tetravalent Ti$^{4+}$ ions in 
Ba$_{1-x}$Bi$_x$Ti$_{1-y}$Fe$_y$O$_3$ has been 
investigated.\cite{dhkim14} 
Similarly as in BaTi$_{1-x}$Fe$_{x}$O$_3$,
the valence states of Fe ions are expected to play a very 
important role in Ba$_{1-x}$Bi$_x$Ti$_{1-y}$Fe$_y$O$_3$. 
Hence, in order to understand the origin of multiferroicity, 
it is crucial to determine the valence states of Fe ions 
directly by using element-specific experimental methods.
In this aspect, soft X-ray absorption spectroscopy (XAS) 
is a powerful experimental tool for studying the valence 
states of TM ions in solids.\cite{Groot90,Laan92}

In this work, we have investigated the electronic structures
of Ba$_{1-x}$Bi$_x$Ti$_{0.9}$Fe$_{0.1}$O$_3$ ($0 \le x \le 0.12$) 
by employing photoemission spectroscopy (PES) and XAS. 
PES provides a direct observation of the occupied part 
of the electronic structures.\cite{Hufner95} 
TM $L$-edge ($2p$) XAS involves the $2p \rightarrow 3d$ 
absorption, and so it is a very powerful experimental tool 
for studying the element-specific valence and spin states 
of TM ions in solids.\cite{Groot90,Laan92} 

\section{Experiments Details}

Polycrystalline Ba$_{1-x}$Bi$_x$Ti$_{0.9}$Fe$_{0.1}$O$_3$ 
($0 \le x \le 0.12$) samples were synthesized by using 
solid-state reaction methods.\cite{Kim14}
PES measurements were performed at the 8A1 beam line,
and XAS measurements were performed at both the 8A1 and 2A 
beam lines of the Pohang Light Source (PLS). 
The base pressure of the XAS/PES chamber was better 
than $3 \times 10^{-10}$ Torr.
XAS and PES spectra were obtained at room temperature.
XAS data were obtained by employing the total electron yield mode
with the photon energy resolution of $\sim 100$ meV
at $h\nu\approx 600$ eV.
The overall instrumental resolution of the PES spectra
was about $\sim 0.4$ eV at $h\nu \sim 600$ eV. 
All the XAS spectra were normalized to the incident photon flux.

\section{Results and discussion}
\begin{figure*}
\includegraphics[angle=270,width=13cm]{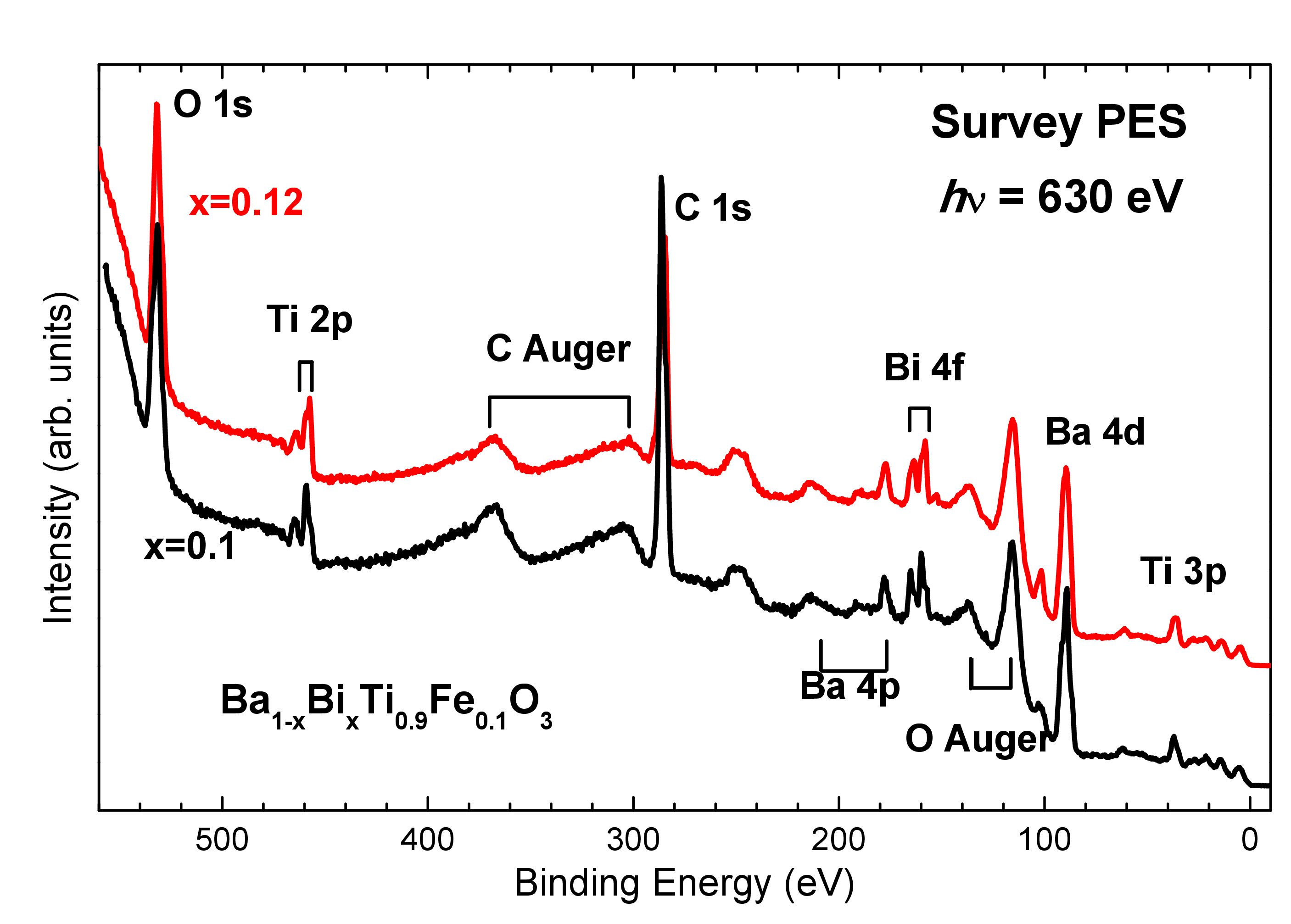}
\caption{(Color online)
        PES survey spectra of 
	Ba$_{1-x}$Bi$_{x}$Ti$_{0.9}$Fe$_{0.1}$O$_3$
	($x$=0.1, 0.12), obtained with $h\nu\approx 630$ eV.}
\label{pes}
\end{figure*}

Figure~\ref{pes} shows the survey PES spectra of 
Ba$_{1-x}$Bi$_{x}$Ti$_{0.9}$Fe$_{0.1}$O$_3$ for $x$=0.1 and 0.12, 
which are shown in the binding energy (BE) scale. 
These PES spectra are obtained with $h\nu\approx 630$ eV.
Here we have chosen to show the samples with high Bi 
concentrations because they can reveal both Ba- and Bi-related 
peaks. All the characteristic core-level peaks of 
Ba$_{1-x}$Bi$_{x}$Ti$_{0.9}$Fe$_{0.1}$O$_3$ are
observed in the measured survey PES spectra, such as 
O $1s$ ($\sim 530$ eV),  Ti $2p$ ($\sim 460$ eV), 
Bi $4f$ ($\sim 160$ eV), and Ba $4d$ ($\sim 90$ eV) core levels. 
The large peak around $\sim 290$ eV is the C $1s$ core-level
peak, which is often observed in polycrystalline samples.
This carbon peak is likely to be due to the impurity carbons 
present in the grain boundaries of polycrystalline samples,
which are not involved in bonding of the solids and do not 
affect the physical properties the samples. This figure shows 
that the Ba$_{1-x}$Bi$_{x}$Ti$_{0.9}$Fe$_{0.1}$O$_3$ 
samples employed in this work are of good quality.

\begin{figure*}
\includegraphics[angle=270,width=15cm]{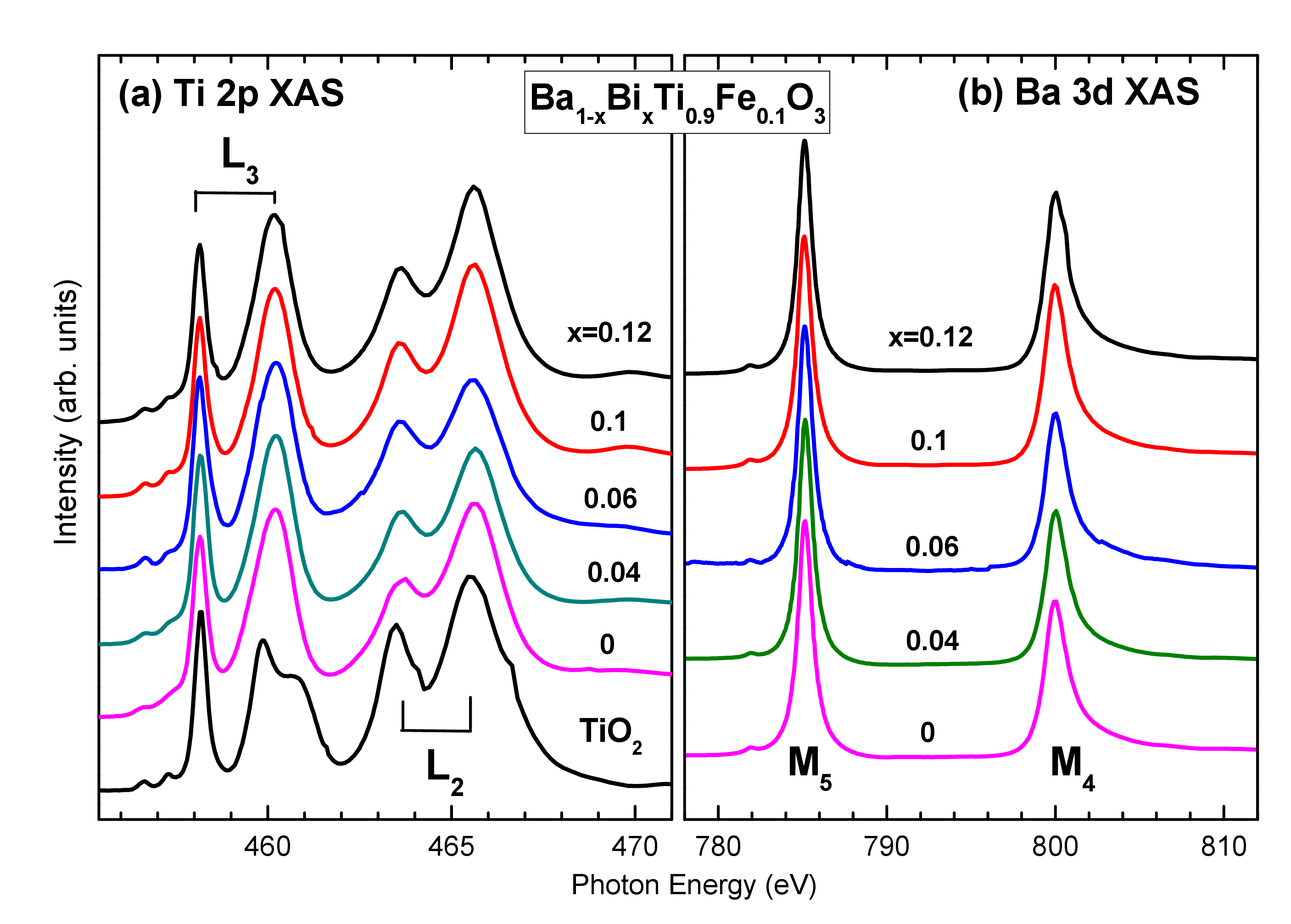}
\caption{(Color online)
(a) Comparison of the Ti $2p$ XAS spectra of 
	Ba$_{1-x}$Bi$_x$Ti$_{0.9}$Fe$_{0.1}$O$_3$ 
	to that of TiO$_2$ ($0 \le x \le 0.12$). 
(b) Comparison of the Ba $3d$ XAS spectra of 
	Ba$_{1-x}$Bi$_x$Ti$_{0.9}$Fe$_{0.1}$O$_3$
	($0 \le x \le 0.12$). }
\label{tiba}
\end{figure*}

Figure~\ref{tiba}(a) shows the measured Ti $2p$ ($L$-edge)
XAS spectra of Ba$_{1-x}$Bi$_{x}$Ti$_{0.9}$Fe$_{0.1}$O$_3$ 
($x$=0, 0.04, 0.06, 0.1, 0.12) in comparison to that of 
tetravalent (Ti$^{4+}$) reference oxide of TiO$_2$
(our data and [Ref.~\cite{Soriano93}]. The first two peaks in the 
low-$h\nu$ region and the other two peaks in the high-$h\nu$ 
region correspond to the Ti $L_3$ and $L_2$ peaks, respectively. 
Here $L_3$ and $L_2$ represent the spin-orbit-split 
$2p_{3/2}$ ($L_3$) and $2p_{1/2}$ ($L_2$) peaks, which arise 
from  the spin-orbit coupling of the Ti $2p$ core hole.
The Ti $2p$ XAS spectra of Ba$_{1-x}$Bi$_{x}$Ti$_{0.9}$Fe$_{0.1}$O$_3$ 
are found to be nearly identical to one another for all $x$, 
indicating that the valence states of Ti ions do not change 
with $x$.  Further, they are very similar to that of TiO$_2$, 
providing evidence that the valence states of Ti ions in
Ba$_{1-x}$Bi$_{x}$Ti$_{0.9}$Fe$_{0.1}$O$_3$ are tetravalent 
(Ti$^{4+}$), having the $3d^0$ ground-state configuration.

Figure~\ref{tiba}(b) shows the measured Ba $3d$ ($M$-edge)
XAS spectra of Ba$_{1-x}$Bi$_{x}$Ti$_{0.9}$Fe$_{0.1}$O$_3$ 
($x$=0, 0.04, 0.06, 0.1, 0.12). Here $M_5$ and $M_4$ represent 
the spin-orbit-split $3d_{5/2}$ and $3d_{3/2}$ peaks. 
This comparison shows that the Ba $3d$ XAS spectra of 
Ba$_{1-x}$Bi$_{x}$Ti$_{0.9}$Fe$_{0.1}$O$_3$ are nearly identical 
to one another for all $x$. This finding indicates that 
the valence states of Ba ions do not change with 
the substitution of Bi ions ($x$).
Since the stable valence states of Ba ions are divalent, 
the findings of Fig.~\ref{tiba}(b) imply that Ba ions are divalent 
(Ba$^{2+}$) in Ba$_{1-x}$Bi$_{x}$Ti$_{0.9}$Fe$_{0.1}$O$_3$ 
for $x \le 0.12$.

\begin{figure*}
\includegraphics[angle=270,width=16cm]{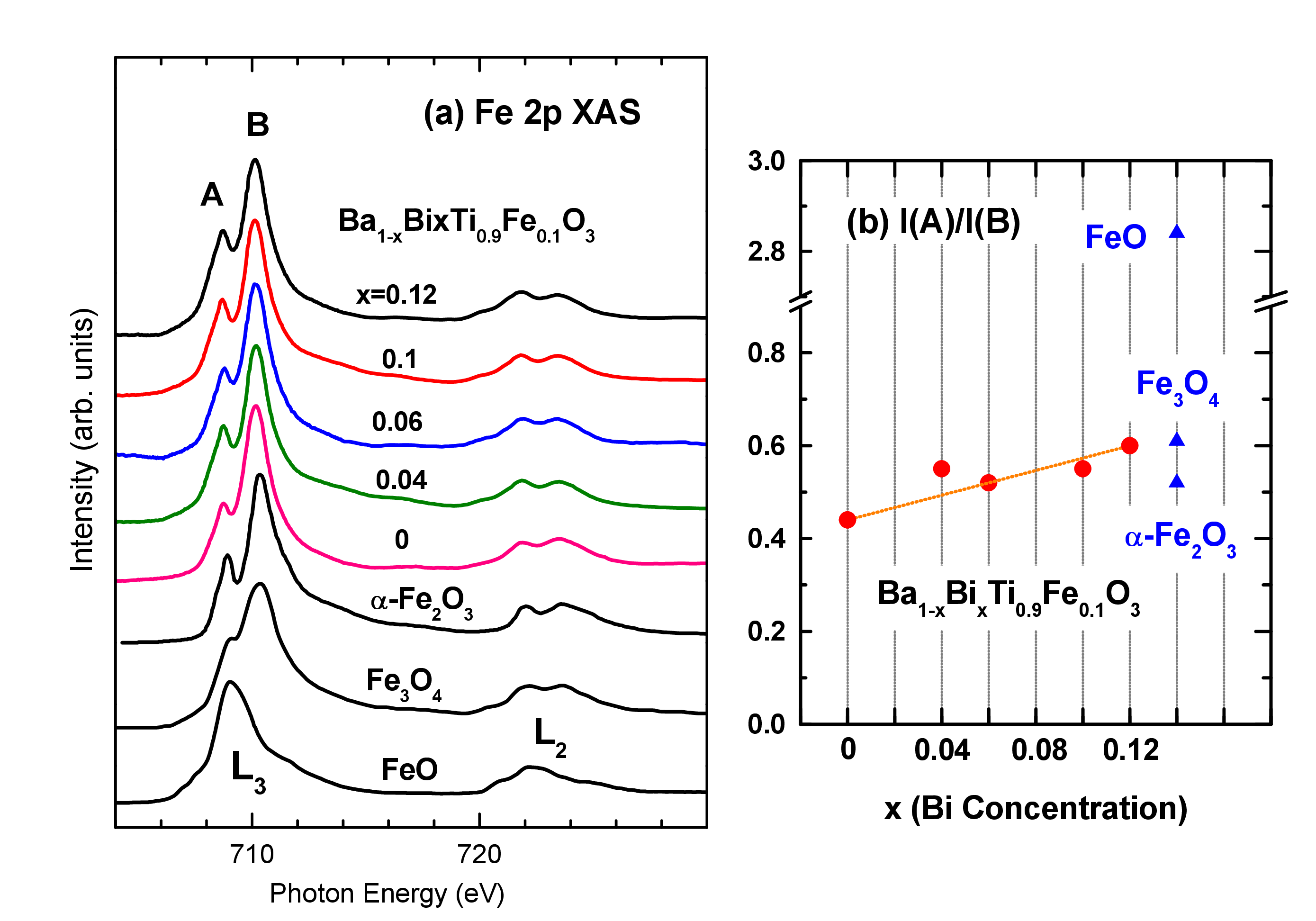}
\caption{(Color online)
        (a) The Fe 2p XAS spectra of 
	Ba$_{1-x}$Bi$_{x}$Ti$_{0.9}$Fe$_{0.1}$O$_3$ 
	($0 \le x \le 0.12$) 
	in comparison with those of reference Fe oxides of 
	trivalent (Fe$^{3+}$) $\alpha$-Fe$_2$O$_3$, 
	divalent (Fe$^{2+}$) FeO, and 
	mixed-valent (Fe$^{2+}$-Fe$^{3+}$) Fe$_3$O$_4$. 
        (b) Plot of the intensity ratio of Fe $L_3$ 
	XAS spectra of Ba$_{1-x}$Bi$_{x}$Ti$_{0.9}$Fe$_{0.1}$O$_3$
	and reference Fe oxides.  }
\label{fe2p}
\end{figure*}

We now discuss on the valence states of Fe ions
in Ba$_{1-x}$Bi$_{x}$Ti$_{0.9}$Fe$_{0.1}$O$_3$.
Figure~\ref{fe2p}(a) shows the measured  Fe $2p$ ($L$-edge)
XAS spectra of Ba$_{1-x}$Bi$_{x}$Ti$_{0.9}$Fe$_{0.1}$O$_3$ 
($0 \le x \le 0.12$).
As a guide of the valence states of Fe ions, they are compared
to those of reference Fe oxides, such as  
divalent (Fe$^{2+}$) FeO (Ref.~\cite{Regan01,Kupper04}), 
trivalent (Fe$^{3+}$) $\alpha$-Fe$_2$O$_3$ 
(Ref.~\cite{Regan01,Kupper04,Kim06}), and
and  mixed-valent (Fe$^{2+}$-Fe$^{3+}$) Fe$_3$O$_4$ 
(Ref.~\cite{Kupper04}).
Similarly as in Ti $2p$ XAS spectra,
$L_3$ and $L_2$ peaks represent the spin-orbit-split 
$L_3$ ($2p_{3/2}$) and $L_2$ ($2p_{1/2}$) peaks.
According to the comparison in Fig.~\ref{fe2p}(a), 
the line shapes of the Fe $2p$ XAS spectra of 
Ba$_{1-x}$Bi$_{x}$Ti$_{0.9}$Fe$_{0.1}$O$_3$ are similar
to that of $\alpha$-Fe$_2$O$_3$ (Fe$^{3+}$)
but quite different from that of FeO (Fe$^{2+}$).
This observation indicates that Fe ions in 
Ba$_{1-x}$Bi$_{x}$Ti$_{0.9}$Fe$_{0.1}$O$_3$ are close to 
being trivalent (Fe$^{3+}$), 

On the other hand, the intensities of the low-$h\nu$ peak 
(peak "$A$") with respect to that of the  high-$h\nu$ peak 
(peak "$B$") in Ba$_{1-x}$Bi$_{x}$Ti$_{0.9}$Fe$_{0.1}$O$_3$
are slightly different from that of $\alpha$-Fe$_2$O$_3$.
Further, I($A$)/I($B$ (=the relative intensity of peak "$A$" 
to that of peak "$B$") changes slightly for different $x$ 
values, which is shown better in the relative intensity 
ratio of I($A$)/I($B$) in Fig.~\ref{fe2p}(b). 
In order to interpret the meaning of I($A$)/I($B$) values
in Ba$_{1-x}$Bi$_{x}$Ti$_{0.9}$Fe$_{0.1}$O$_3$, we examine 
the lineshapes and I($A$)/I($B$) of reference Fe oxides.
Note that, in divalent (Fe$^{2+}$) FeO, the peak "$A$" is 
dominant and the peak "$B$" appears as a high-$h\nu$ shoulder.
I($A$)/I($B$) increases from trivalent (Fe$^{3+}$) 
$\alpha$-Fe$_2$O$_3$ (0.52) to  mixed-valent (Fe$^{2+}$-Fe$^{3+}$) 
Fe$_3$O$_4$ (0.63) and divalent (Fe$^{2+}$) FeO ($>1$). 
This trend is shown more clearly in Fig.~\ref{fe2p}(b). 

Figure~\ref{fe2p}(b) shows that I($A$)/I($B$)  
in Ba$_{1-x}$Bi$_{x}$Ti$_{0.9}$Fe$_{0.1}$O$_3$ increases 
slightly with increasing $x$. I($A$)/I($B$) for $x$=0 is lower 
than that of $\alpha$-Fe$_2$O$_3$, but I($A$)/I($B$) 
for $x$=0.12 is similar to that of Fe$_3$O$_4$. 
Such a trend in Ba$_{1-x}$Bi$_{x}$Ti$_{0.9}$Fe$_{0.1}$O$_3$ 
and the relative magnitudes with the reference Fe oxides  
indicate that, in $x$=0 (BaTi$_{0.9}$Fe$_{0.1}$O$_3$), 
Fe ions are nearly trivalent (Fe$^{3+}$), and that, 
with the substitution of Bi ions for Ba ions ($x>0$), 
the Fe$^{2+}$ component appears, resulting in  
Fe$^{2+}$-Fe$^{3+}$ mixed-valent states.
For $x$=0.12 (Ba$_{0.88}$Bi$_{0.12}$Ti$_{0.9}$Fe$_{0.1}$O$_3$),
the valence states of Fe ions become similar to that
of Fe$_3$O$_4$ ($v$(Fe)$\sim 2.67$).
This finding is consistent with the expected trend
since Bi ions are generally trivalent (Bi$^{3+}$) 
whereas Ba ions are divalent (Ba$^{2+}$), as confirmed by
the measured Ba $3d$ XAS spectra (see Fig.~\ref{tiba}(b)).
Considering the charge neutrality conditions in insulating 
oxides, the substitution of trivalent Bi$^{3+}$ ions
for divalent Ba$^{2+}$ ions would cause the decreasing
valence states of Fe ions.    

Our Fe $2p$, Ti $2p$, and Ba $3d$ XAS study on 
Ba$_{1-x}$Bi$_{x}$Ti$_{0.9}$Fe$_{0.1}$O$_3$ 
shows that Ti ions are tetravalent (Ti$^{4+}$) 
and Ba ions are divalent (Ba$^{2+}$) for all $x$
in Ba$_{1-x}$Bi$_{x}$Ti$_{0.9}$Fe$_{0.1}$O$_3$ 
($x \le 0.12$). In contrast, with increasing $x$,
the valence states of Fe ions decrease
from being nearly trivalent ($v$(Fe)$\sim 3$) for $x$=0
(no Bi ions) to being mixed-valent for $x>0$
(Fe$^{2+}$-Fe$^{3+}$ mixed-valent)
and to become $v$(Fe)$\sim 2.6$ for $x$=0.12.
This finding refutes the suggestion of the  
Fe$^{3+}$-Fe$^{4+}$ mixed-valence states in
Ba$_{1-x}$Bi$_x$Ti$_{0.9}$Fe$_{0.1}$O$_3$. 
Hence the idea of the exchange interactions among 
Fe$^{3+}$-Fe$^{4+}$, Fe$^{3+}$-Fe$^{3+}$, 
and Fe$^{4+}$-Fe$^{4+}$, need be examined more carefully. 
This work suggests that the valence states of Fe
ions play an important role in determining 
the electronic and magnetic properties of 
Ba$_{1-x}$Bi$_x$Ti$_{0.9}$Fe$_{0.1}$O$_3$. 

\section{CONCLUSIONS}

The electronic structures of of multiferroic oxides of
Ba$_{1-x}$Bi$_x$Ti$_{0.9}$Fe$_{0.1}$O$_3$ ($0 \le x \le 0.12$). 
have been investigated by employing synchrotron-radiation 
excited PES and XAS. 
Via Fe and Ti $2p$ XAS measurements, the valence states of 
Fe and Ti ions have been determined experimentally.  
The valence states of Fe ions are found to be
Fe$^{2+}$-Fe$^{3+}$ mixed-valent for $x>0$
but nearly trivalent for $x$=0.
The valence states of Fe ions are found to decreases
from being nearly trivalent for $x$=0 ($v$(Fe)$\sim 3$) 
to $v$(Fe)$\sim 2.6$ for $x$=0.12.
The valence states of Ti ions do not change with $x$ 
for $x \le 0.12$, and stay being tetravalent (Ti$^{4+}$). 
The valence states of Ba ions are close to being divalent
(Ba$^{2+}$) and remain unchanged for $x \le 0.12$. 
The decreasing trend of valence states of Fe ions 
in Ba$_{1-x}$Bi$_x$Ti$_{0.9}$Fe$_{0.1}$O$_3$ with $x$
will play an important role in the electronic 
and magnetic properties of this system.

\begin{acknowledgments}
This work was supported by the NRF under Contract No. 2014R1A1A2056546, 
and in part by the Research Fund, 2015 of 
the Catholic University of Korea.
Experiments at PLS were supported by MSIP and PAL.

\end{acknowledgments}

\end{document}